\documentstyle[preprint,aps]{revtex}

\newcommand{\etal}{{\em et al.}}
\newcommand{\hel}{${}^3$He}
\newcommand{\cdeg}{${}^\circ$}

\begin{document}
\draft
\title{
Inclusive Scattering of Polarized Electrons on Polarized $^{{\boldmath 3}}$He: Effects
 of Final State Interaction
and the Magnetic Form Factor of the Neutron
}

\author{S. Ishikawa}
\address{Department of Physics, Hosei University,
Fujimi 2-17-1, Chiyoda, Tokyo 102, Japan}
\author{J. Golak and H. Wita\l a}
\address{Institute of Physics, Jagellonian University, 
PL-30059 Cracow, Poland}
\author{H. Kamada\footnote{present address: Institut f\"ur
    Kernphysik, Fachbereich 5 der Technischen Hochschule Darmstadt, D-64289
    Darmstadt, Germany   }, W. Gl\"ockle  and 
    D. H\"uber\footnote{present address: Los Alamos National Laboratory, Theoretical 
    Division, Los Alamos, NM 87545 } }
\address{Institut f\"ur Theoretische Physik II, 
Ruhr-Universit\"at Bochum, D-44780 Bochum, Germany}

\date{\today}
\maketitle

\begin{abstract}
Effects of final state interaction on asymmetries in inclusive 
scattering of polarized electrons on polarized  \hel{}
are investigated using consistent \hel{} bound state wave function 
and 3N continuum scattering states. Significant effects are found, 
which influence the extraction of the magnetic neutron form factor from $A_{T'}$.
The enhancement found experimentally for $A_{TL'}$ near the 3N breakup
threshold, which could not be explained in calculations
carried through in plane wave impulse approximation up to now,
occurs now also in theory if the full final state interaction is included.
\end{abstract}

\pacs{13.40.Gp, 13.60.Hb, 21.45.+v, 24.70.+s, 25.30.Fj}

\narrowtext

\section{Introduction}
The electromagnetic form factor of the nucleons are of fundamental
interest in nuclear and particle physics.While the proton form factors
have been determined from elastic electron-proton scattering over
a wide range of momentum transfers with good accuracy~\cite{cite1}, this is
not the case for the neutron, since no free neutron targets exist.
One is therefore forced to extract information on the neutron from electron
scattering on light nuclei. Obviously ambiguities arising from nuclear
structure and reaction mechanisms should be minimized. So far mainly the
deuteron has been used as a target~\cite{cite2}. 
The $^3$He nucleus has also attracted much attention
as an ideal target~\cite{cite3,cite4}. If one assumes that the $^3$He wave
function is spatially symmetric (antisymmetric in spin-isospin space), then the spins of the
two protons are coupled to zero and the spin of $^3$He is carried by the
neutron alone.
Under this simplifying assumption a polarised $^3$He nucleus can be considered 
to be a polarised neutron.
Now this picture of the $^3$He wave function, the so-called principal S-state
approximation, is valid to about 92~\% with respect to its norm.
( This referes to Bonn B potential \cite{cite19}, which we use in this article )
Motivated by that attractive feature, recently several experiments 
have been performed, where longitudinally polarized electrons with
helicities h(=$\pm$1) have been inclusively scattered on polarized $^3$He
targets~\cite{cite5,cite6,cite7,cite8,cite9}.
The aim was to measure the asymmetries

\begin{equation}
A = 
\frac{\sigma(h=+1) - \sigma(h=-1)}{\sigma(h=+1) + \sigma(h=-1)}
,
\label{eq1}
\end{equation}

\noindent
depending on the spin direction of $^3$He. These asymmetries are expected
to be sensitive to the electromagnetic form factors of the neutron. The
data have been analysed so far in plane wave impulse 
approximation~\cite{cite10,cite11}
and based on a single nucleon current operator. That approximation neglects
the interaction between the nucleon which absorbed the photon and the two 
other nucleons.
 
It is the aim of this investigation to remove that theoretical uncertainty
and to treat the $^3$He bound state wave function and the 3N continuum
representing the final 3N scattering state on an equal footing, using exact solutions
of three-body Faddeev equations
based on realistic NN forces. Our theoretical formalism is described 
in section II and our results in comparison to the data in section III.
A Summary is given in section IV.

\section{Theory}

In recent articles~\cite{cite12,cite13,cite14,cite15,cite16} 
we studied elastic and inelastic electron 
scattering
on $^3$He corresponding to unpolarized experiments. So far our dynamical
picture is: a nonrelativistic framework, a single nucleon current operator
and the exact treatment of realistic NN forces among the three nucleons.
For the relatively low momentum transfers considered up to now that
picture was quite successful and the final state interaction (FSI) among the
three nucleons played a significant role. Now we apply that dynamical 
picture to the
scattering of polarized electrons on polarized $^3$He targets under
inclusive conditions. The derivation of the corresponding cross 
section is known~\cite{cite17}. However to stay in line with
the notation in our previous articles and to show its extensions
we just mention the new ingredients.
In the evaluation of the cross section the fixed
electron polarization in the initial state leads to an additional term
proportional to the helicity h on top of the usual expression for the
electron tensor

\begin{eqnarray}
L ^{\mu \nu}  &\equiv& 
\sum_{s'}  \bar u (k's') \gamma ^{\mu} u(ks) ( \bar u (k's') \gamma ^{\nu}
u(ks) )^{*} \cr
&=& {1\over {2m^{2}} } ( k^{\mu} {k'} ^{\nu} + {k'}^{\mu} k^{\nu} - g^{\mu
  \nu}k \cdot k' + i h \epsilon ^{\mu \nu \alpha \beta} k_{\alpha} k'
_{\beta} ) 
\label{eq2}
\end{eqnarray}

That additional last term in Eq.~(\ref{eq2}) has been evaluated under the
condition, that the electron mass $m$ can be neglected in relation  to its
energy. Straightforward contraction with the hadronic tensor yields the
inclusive  cross section in the lab system

\begin{eqnarray} 
{{d \sigma} \over {d \hat k ' d k_{0}}  } = \sigma _{Mott} \left[ v_{L } R_{L} +
v_{T} R_{T} + h ( v_{T'} R_{T'} + v_{TL'} R_{TL'} ) \right] 
\label{eq3}
\end{eqnarray}

The unprimed terms are the familiar ones for the unpolarized 
set up~\cite{cite16}. The primed terms are: kinematical factors 
from the electron tensor 

\begin{eqnarray}
v_{T'} = \sqrt{{-Q^{2} \over {\vec Q ^{2}} } + \tan^{2} {\Theta \over 2}}
\tan {\Theta \over 2} 
\label{eq4}
\end{eqnarray}

\begin{eqnarray}
v_{TL'} = {1 \over \sqrt{2} } {Q^{2} \over {\vec Q ^{2}} }  
\tan {\Theta \over 2} 
\label{eq5}
\end{eqnarray}

and structure functions related to the hadronic tensor 

\begin{eqnarray}
R_{T'} = \sum_{m' \, {\tau}' } \int df' \delta (M+\omega -P_{0}' ) ( \vert N_{1 } \vert
^{2} - \vert N_{-1}\vert ^{2} ) 
\label{eq6}
\end{eqnarray}

\begin{eqnarray}
R_{TL'} = - \sum_{m' \, {\tau}' } \int df' \delta ( M + \omega -P_{0 } ' ) 2
\mbox{Re}\left[ N_{0} ~( N_{1} + N_{-1} ) ^{*} \right]
\label{eq7}
\end{eqnarray}

Here $Q= ( \omega $, $\vec Q $ ) is the four momentum of the photon, $\Theta $
the electron scattering angle, M the target mass and $P_{0} '$ the total
energy of the final state. The summation over all spin and isospin 
magnetic quantum numbers
and momenta in the final state is indicated by $m'$, $ {\tau}'$  and $df'$. 
The nuclear matrix elements $N_{0}$ and $N_{\pm 1 } $ are

\begin{eqnarray}
N_{0} = \langle \Psi_{f'm' {\tau}' } ^{(-)} \vert \rho (\vec Q) \vert \Psi _{^{3}\mbox{He}} \rangle  
\label{eq8}
\end{eqnarray}
\begin{eqnarray}
N_{\pm 1} = \langle \Psi_{f'm' {\tau}' } ^{(-)} \vert j_{\pm 1 } (\vec Q) \vert \Psi
_{^{3}\mbox{He}} \rangle {\rm ,} 
\label{eq9}
\end{eqnarray}
where $ \vert \Psi _{^{3}\mbox{He}} \rangle $ 
is the $^{3}$He ground state, $ \vert \Psi _{f'm' {\tau}' } ^{(-)}\rangle $  
a 3N scattering state with the asymptotic quantum
numbers $f'm' {\tau}' $, $\rho (\vec Q ) $ the electromagnetic hadronic density 
operator and
$j_{\pm 1 }(\vec Q)$ the spherical components of the electromagnetic hadronic
current operator.
Since we use a nonrelativistic framework, the argument of the $\delta$-function
in Eqs.~(\ref{eq6}-\ref{eq7}) is 

\begin{eqnarray}
M+ \omega -P_{0} ' &=& 
\epsilon _{^{3}
\mbox{He}} + \omega -{ {\vec Q ^{2}} \over
  {6 m_{N}}} - E_{f'} \cr
&\equiv& E-E_{f'}
\label{eq10}
\end{eqnarray}

\noindent
where $\epsilon_{^{3}\mbox{He}}$ is the $^{3}$He binding energy (negative),
$m_{N}$ the nucleon mass, the final total momentum $\vec P ' = \vec Q$ and
$E_{f'}$ the internal 3N energy related to the quantum numbers $f'$.

In evaluating the primed structure functions we can generalise a method
proposed in~\cite{cite14}. Let us define 

\begin{eqnarray}
{\cal R}
_{AB} &\equiv& \sum_{m' {\tau}' } \int df' \delta ( E-E_{f'}) \langle \Psi ^{(-)}
_{f'm'\tau '} \vert A \vert \Psi_{^{3}\mbox{He}} \rangle \langle \Psi _{f'm'\tau'} ^{(-)
  } \vert B \vert \Psi _{^{3}\mbox{He}} \rangle ^{*} \cr
&=& 
\sum_{m' {\tau}' } \int df' \langle \Psi_{^{3}\mbox{He} }  \vert B ^{\dagger} \vert \Psi_{f'm'\tau'}
^{(-)} \rangle \delta ( E-E_{f'}) \langle \Psi _{f'm'\tau'} ^{(-)} \vert A \vert
\Psi _{^{3}\mbox{He}} \rangle \cr
&=& 
\sum_{m' {\tau}' } \int df' \langle \Psi _{^{3}\mbox{He} } \vert B ^{\dagger} \delta ( E-H )
\vert  \Psi _{f'm'\tau'} ^{(-)}  \rangle  \langle \Psi _{f'm'\tau'} ^{(-)} \vert A \vert
\Psi _{^{3}\mbox{He}} \rangle \cr
&=&
\langle \Psi _{^{3 } \mbox{He}} \vert B ^{\dagger} \delta ( E - H ) A \vert
\Psi _{^{3}\mbox{He}} \rangle 
\label{eq11}
\end{eqnarray}

\noindent
We introduced the 3N Hamiltonian $H$ and used the completeness relation (the
ground state does not contribute, since $E$ lies in the 3N continuum). 
In our case the operators $A$ and $B$ are either $\rho (\vec Q)$ or 
$j_{\pm} (\vec Q) $.

If the operators $A$ and $B$ are different the method proposed in~\cite{cite14} to
evaluate the last expression in Eq.~(\ref{eq11}) has to be generalised  to 

\begin{eqnarray}
{\cal R}
_{AB} &=& { 1 \over {2 \pi i  }} \langle \Psi _{^{3}\mbox{He}} \vert 
B^{\dagger}{ 1 \over
  { E- i \epsilon - H  }} A \vert \Psi _{^{3}\mbox{He}} \rangle 
- { 1 \over { 2 \pi i }} \langle \Psi _{^{3 } \mbox{He }} \vert B ^{\dagger }
{ 1 \over { E + i \epsilon - H }} A \vert \Psi _{^{3}\mbox{He}} \rangle 
\cr 
&\equiv& 
{ 1 \over { 2 \pi i } } \langle \Psi _{^{3 } \mbox{He} } \vert B ^{\dagger }
\vert \Psi _{A} ^{(-)} \rangle - { 1 \over { 2 \pi i }} \langle \Psi
_{^{3}\mbox{He}} \vert B ^{\dagger } \vert \Psi _{A} ^{(+)} \rangle 
\label{eq12}
\end{eqnarray}

\noindent
We introduced 

\begin{eqnarray}
\vert \Psi _{A } ^{(\pm)} \rangle 
\equiv { 1 \over { E  \mp i \epsilon - H } } A
\vert \Psi _{^{3 } \mbox{He} } \rangle 
\label{eq13}
\end{eqnarray}

Now 

\begin{eqnarray} 
\langle \Psi _{^{3 } \mbox{He}} \vert B ^{\dagger} \vert \Psi _{A } ^{(-)}
\rangle &=& \langle  \Psi ^{(-)} _{A} \vert B \vert \Psi _{^{3}\mbox{He}} 
\rangle ^{*} 
\cr
&=&
\langle \Psi _{^{3}\mbox{He}} \vert A ^{\dagger} { 1 \over { E + i \epsilon -H
    } } B \vert \Psi _{^{3}\mbox{He}} \rangle ^{*} 
\cr
&\equiv&
\langle \Psi _{^{3}\mbox{He}} \vert A ^{\dagger } \vert \Psi _{B} ^{(+)} 
\rangle {}^{*} 
\label{eq14}
\end{eqnarray}

with 

\begin{eqnarray}
\vert \Psi _{B}^{(+)} \rangle = { 1 \over { E + i \epsilon - H } } B \vert
\Psi _{^{3} \mbox{He}} \rangle 
\label{eq15}
\end{eqnarray} 

Therefore 
we get 

\begin{eqnarray}
{\cal R}
_{AB} = { 1 \over { 2 \pi i } } ( \langle \Psi _{^{3} \mbox{He} } \vert
A^{\dagger} \vert \Psi ^{(+) } _{B } \rangle^{*} - \langle \Psi
_{^{3}\mbox{He} } \vert B^{\dagger } \vert \Psi _{A } ^{(+)} \rangle )
\label{eq16}
\end{eqnarray}

For $A=B$ we recover the old result~\cite{cite14}

\begin{eqnarray} 
{\cal R}
_{AA} = - { 1\over \pi } \mbox{Im} \langle \Psi _{^{3}\mbox{He}} \vert
  A^{\dagger } \vert \Psi _{A}^{ (+)} \rangle 
\label{eq17}
\end{eqnarray}

The states $\vert \Psi _{A,B} ^{(+)} \rangle$, defined in Eqs.~(\ref{eq13}) and
(\ref{eq15}) contain all the complexity of the interaction among the three
nucleons and are evaluated as in ~\cite{cite14,cite16} using the Faddeev scheme. We get
\begin{eqnarray}
\Psi _{C} ^{(+)}= G_{0 } ( 1 + P) U_{C}
\label{eq18}
\end{eqnarray}
with 
\begin{eqnarray}
U_{C} =  ( 1 + tG_{0} ) C^{(1)} \vert \Psi _{^{3}\mbox{He}} \rangle + tG_{0} P U_{C} 
\label{eq19}
\end{eqnarray}
Here  $C$ is either $A$ or $B$ (for instance $\rho$ or $j_\pm$) and we assumed that $A$ or $B$ can 
be decomposed as 

\begin{eqnarray}
C = \sum _{i=1} ^{3} C ^{(i)} {\rm .} 
\label{eq20}
\end{eqnarray}

Further $t$ 
is the NN t-matrix, $G_{0}$ the free 3N propagator and $P$ the sum of a cyclic
 and anticyclic permutation of 3 objects.

The Faddeev equation (\ref{eq19}) has been introduced and handled numerically
before in \cite{cite16}. 

Inserting Eq.~(\ref{eq19}) into Eq.~(\ref{eq16}) we get
\begin{eqnarray}
{\cal 
 R}_{AB} &=& { 1 \over { 2 \pi i }} \left( \langle 
\Psi _{^{3 }\mbox{He}} \vert A^{\dagger} G_{0 } ( 1+ P ) U_{B} \rangle ^{*} -
\langle \Psi _{^{3}\mbox{He}} \vert B^{ \dagger} G_{0 } ( 1 + P  ) U_{A}
\rangle \right) 
\cr
&=&
{3 \over {2 \pi i }} \left( \langle \Psi _{^{3}\mbox{He}} \vert A^{(1)\dagger
    }  G_{0 }  ( 1 + P) U_{B} \rangle ^{*} - 
\langle \Psi _{^{3} \mbox{He}} \vert B^{(1)\dagger } G_{0 } ( 1 + P )
U_{A} \rangle \right) 
\label{eq21}
\end{eqnarray}

In the last step we used Eq.~(\ref{eq20}) and the fact that the states to the left
and right of $A^{\dagger}$ or $B^{\dagger}$ are antisymmetrical.

Regarding Eqs.~(\ref{eq6}) and (\ref{eq7}) we see that the expressions ${\cal R}_{AB}$ are either of the
form ${\cal R}_{AA}$ and therefore real or for $A\ne B$ one has to take the real part
thereof. Thus in general we have to add the step

\begin{eqnarray}
R_{AB} &=& 
\mbox{Re} \left[ {\cal R}_{AB} \right] 
\cr
&=&
{ 3 \over { 2 \pi}} \mbox{ Im} \left[
 \langle \Psi_{^{3\mbox{He}}}\vert A^{(1) \dagger} G_{0 } ( 1 +
P) U_{B} \rangle ^{* } -
\langle \Psi _{^{3 } \mbox{He}} \vert B ^{(1) \dagger } G_{0} ( 1 + P ) U_{A}
\rangle \right]  {\rm .}
\label{eq22}
\end{eqnarray}
This applies to $R_{TL'}$ in our case.

Further considerations require a partial wave decomposition and taking the
polarization  of $^{3} $He into account. 
We introduce our standard  basis  in momentum space~\cite{cite18} 
\begin{eqnarray}
\vert p q \alpha \rangle = \vert p  q ( l s ) j ( \lambda {1 \over 2} ) J
{\cal J} M ( t { 1\over 2} ) T M_{T} \rangle {\rm ,}
\label{eq23}
\end{eqnarray}
where p and q are magnitudes  of Jacobi momenta and the set of discrete quantum
numbers $\alpha$ comprises angular momenta, spins and isospins for a three-nucleon
system.
The $^{3}$He state polarized in the direction $\theta^{*}$, $\phi^{*}$ is 
\begin{eqnarray}
\vert \Psi _{^{3}\mbox{He}} m \rangle _{\theta^{*} \phi^{*}}  = \sum _{ m'} \vert \Psi
_{^{3}\mbox{He}} m' \rangle 
 D ^{(1/2)} _{m',m} ( -\phi ^{*}, -\theta ^{*}, 0  ) {\rm ,}
\label{eq24}
\end{eqnarray}  
where $\vert \Psi _{^{3}\mbox{He}}m \rangle  $ is quantised with respect to the
z-direction and the Wigner D-function occurs as
\begin{eqnarray}
D^{(1/2)} _{m',m} ( -\phi ^{*}, -\theta ^{*},0 ) = e^{-im'\phi^{*}} 
\left( \matrix{ \cos{ \theta ^{*} \over 2 } & -\sin{ \theta^{*} \over 2 } \cr 
\sin{ \theta  ^{*} \over 2 } & \cos{ \theta^{*} \over 2 }  \cr } 
\right)
\label{eq25}
\end{eqnarray}
Using all that we get 

\[
{}_{\theta^{*}\phi ^{*} } \langle \Psi _{^{3}\mbox{He}}  m \vert
 B ^{(1) \dagger } G_{0 }
( 1 + P ) U_{A} \rangle 
= 
\sum _{\alpha} \int _{0 }^{ \infty } p^{2 } dp \int _{0 } ^{\infty} q^{2 }dq
{ 1 \over { E + i \epsilon - { p^{2 } \over m } - { 3 \over {4 m}} q^{2} }} 
\]
\begin{eqnarray}
\sum _{m'} \sum _{m''} 
D_{m',m} ^{{(1/2)}* } D^{(1/2)} _{m'',m} 
\langle p q \alpha \vert ( 1 + P )  B ^{(1)} \vert \Psi _{^{3}\mbox{He}} {m'}
\rangle ^{*} 
\langle p q \alpha \vert U_{A} m''\rangle 
\label{eq26}
\end{eqnarray}

Note that the state $\vert U_{A} \rangle $ depends on the magnetic quantum
number $m$ of $^{3}$He through the driving term in Eq.(\ref{eq19}).

Let us illustrate how the dependence on the magnetic quantum number $m$ of the
$^{3}$He polarization enters into the four structure functions of Eq.~(\ref{eq3})
in two examples. The remaining ones are worked out in the Appendix.
The longitudinal structure function $R_{L}$ has the form 
\begin{eqnarray}
R_{L} &=& - {3 \over \pi } \mbox{Im} \left[ \langle \Psi _{{^{3}\mbox{He}}} \vert
\rho ^{(1) \dagger}  G_{0 } (1 + P ) U_{\rho} \rangle  \right]
\cr
&{}&
\cr
&=&
-{ {3 \over \pi }} \mbox{Im } 
[ 
{\sum \hskip -13 pt \int}
{ 1 \over { E + i \epsilon - {
      p^{2 }\over m } - { 3 \over {4 m } } q^{2 } } } \sum _{m'} \sum _{m''}
\cr
&{}&
D^{(1/2) *} _{m',m}  D^{(1/2)} _{m'',m} \langle p q \alpha \vert ( 1 + P ) \rho
^{(1)}  \vert \Psi _{^{3}\mbox{He }}  m' \rangle ^{* } 
\langle p q \alpha \vert U_{\rho } m''\rangle ] 
\label{eq27}
\end{eqnarray}

The sums in ${\displaystyle {\sum \hskip -13 pt \int}} $ include the summation over the magnetic quantum
number $M$ of the total 3N angular momentum ${\cal J}$. 
We indicate that dependence on $ {\cal J} M$ now explicitely and consider the expression 
\begin{eqnarray}
\sum_{M} \sum_{m'} \sum _{m''} D^{(1/2) *} _{m', m } D^{(1/2)} _{m'',m} 
\langle pq \alpha  {\cal J}  M \vert  ( 1 + P ) \rho ^{(1)} \vert \Psi
_{^{3}\mbox{He}} m' \rangle ^{* } \langle p q \alpha  {\cal J}
 M \vert U_{\rho } m''
\rangle 
\label{eq28}
\end{eqnarray} 
Since we choose the z-axis to lie in the direction $\hat Q$ of the virtual
photon and therefore the density operator $\rho ^{(1)}$ conserves the 3N
magnetic quantum number~\cite{cite16}, 
one has $M=m'=m''$ and the expression Eq.~(\ref{eq28}) simplifies to 
\begin{eqnarray}
&{}&
D^{(1/2) *} _{{1\over 2}, m } D ^{(1/2) } _{{1\over 2},m } \langle p q \alpha
{\cal J}{ 1 \over 2} \vert ( 1  + P ) \rho ^{ (1)} \vert \Psi _{^{3}\mbox{He}}
{1 \over 2}
\rangle \langle p q \alpha {\cal J} {1\over 2} 
\vert U_{\rho } { 1 \over 2 } \rangle 
\cr
&+& 
D^{(1/2) *} _{-{ 1 \over 2},m} D ^{(1/2)} _{-{1 \over 2 }, m } \langle p q
\alpha {\cal J}
 -{1\over 2} \vert (1 + P) \rho ^{(1)} \vert \Psi_{^{3}\mbox{He}} -{1
  \over 2 } \rangle \langle p q \alpha {\cal J}
 -{1 \over 2} \vert U _{\rho} -{ 1
  \over 2 } \rangle 
\label{eq29}
\end{eqnarray}
We used the fact that the $\rho ^{(1)}$ matrix element is real.

Now a detailed look into the partial wave decomposed forms \cite{cite16} reveals the
following symmetry properties
\begin{eqnarray}
\langle p q \alpha {\cal J} -{1 \over 2} \vert ( 1 + P ) \rho ^{(1)} \vert \Psi
_{^{3} \mbox{He}} -{1 \over 2} \rangle 
= (-1) ^{{\cal J} - { 1 \over 2 } } \ \Pi \  \langle p q \alpha {\cal J} 
{1 \over 2 }
\vert ( 1 + P) \rho ^{(1)}
\vert \Psi _{^{3}\mbox{He} } { 1 \over 2} \rangle 
\label{eq30}
\end{eqnarray}
\begin{eqnarray}
\langle p q \alpha {\cal J} -{1 \over 2} \vert U_{\rho} -{1 \over 2} \rangle 
= (-1) ^{{\cal J} - { 1 \over 2 } } \ \Pi \ \langle p q \alpha {\cal J} 
{1 \over 2 }
\vert U_{\rho } { 1 \over 2} \rangle {\rm ,}
\label{eq31}
\end{eqnarray}
where $\Pi$ is the parity of the state $ \vert p q \alpha \rangle $. 
With the help of Eqs.~(\ref{eq30}-\ref{eq31}) it is obvious that 
in Eq.~(\ref{eq29}) the D-functions
can be separated  into the sum  
\begin{eqnarray}
\vert D^{(1/2)} _{{1\over 2}, m  }  \vert ^{2 } +
 \vert D^{(1/2)} _{-{1 \over 2}, m } \vert ^{2 } = 1 
\label{eq32}
\end{eqnarray}
and we end up with
\begin{eqnarray}
R_{L} = - { 3 \over \pi } \mbox{Im} \left[ 
{\sum \hskip -13 pt \int}
{ 1 \over { E + i \epsilon -{
      p^{2 } \over m } - { 3 \over {4 m } } q^{2 }  } } 
\langle p q \alpha {\cal J}
 { 1 \over 2} \vert ( 1 + P ) \rho ^{ (1)} \vert \Psi
_{^{3}\mbox{He}} { 1 \over 2} \rangle \langle p q \alpha {\cal J}
{1\over 2} \vert
U_{\rho } {1 \over 2} \rangle \right] {\rm ,}
\label{eq33}
\end{eqnarray}
which is independent of $m$, the polarization of $^{3}$He.

A corresponding study carried through in the Appendix leads to 
\begin{eqnarray}
R_{T} &=& -{ 3 \over \pi } \mbox{Im} [ 
{\sum \hskip -13 pt \int}
{ 1 \over { E + i \epsilon - {
      p^{2 } \over m} - { 3 \over { 4 m} } q^{2}  } }  
( \langle p q \alpha {\cal J} { 1 \over 2 } \vert ( 1 + P) j_{1} ^{(1)} \vert
  \Psi_{^{3}\mbox{He}} - { 1 \over 2 } \rangle ^{*} \langle p q \alpha {\cal
    J}
 {1
    \over 2 } \vert U _{j_{1}} -{ 1 \over 2} \rangle 
\cr
&+& \langle p q \alpha {\cal J} { 3 \over 2} \vert ( 1 + P) j_{1 }^{(1)} \vert
\Psi_{^{3}\mbox{He}} { 1 \over 2} \rangle ^{*} \langle p q \alpha {\cal J}
 { 3 \over
  2 } \vert U_{j_{1}} { 1 \over 2} \rangle ) ]
\label{eq34}
\end{eqnarray}
Here $j_{1}^{(1)} $ is the spherical +1 component of the current
operator. Again we see that $R_{T}$ is independent of the $^{3}$He target
polarization. 

As the second illustration we regard $R_{TL'}$.
According to Eqs.~(\ref{eq7}), (\ref{eq22}) and (\ref{eq26}) it has the form
\begin{eqnarray}
R_{TL'} &=& - 2 { 3 \over {2 \pi } } \mbox{Im}
\Bigl[ \langle \Psi _{^{3}\mbox{He}} \vert \rho ^{(1) \dagger} G_{0}(1 +P)
U_{j_{1}} \rangle ^{*} -\langle \Psi _{^{3}\mbox{He}} \vert j_{1} ^{(1) 
\dagger} G_{0}(1 +P)
U_{\rho} \rangle 
\cr
&{}&
+\langle \Psi _{^{3}\mbox{He}} \vert \rho ^{(1) \dagger} G_{0}(1 +P)
U_{j_{-1}} \rangle^{*} - \langle \Psi _{^{3}\mbox{He}} \vert j_{-1} ^{(1)
  \dagger} G_{0}(1 +P) 
U_{\rho} \rangle  ]
\cr
&=& - {3 \over \pi} \mbox{Im} [ {\sum \hskip -13 pt \int} \, {1 \over { E- i \epsilon -{p^{2 } \over m}
    -{3\over {4m}} q^{2}  }} \sum_{m',m''} D_{m',m} ^{(1/2) } D_{m''
,m}^{(1/2)* }
\langle p q \alpha \vert (1 + P) \rho ^{(1)} \vert \Psi _{^{3}\mbox{He} } m'
\rangle 
\cr
&{}&
\left( \langle p q \alpha \vert U_{j_{1}} m'' \rangle ^{* } + \langle p
  q \alpha \vert U_{j_{-1}} m'' \rangle ^{*} \right) 
\cr
&{}&-
 {1 \over { E+ i \epsilon -{p^{2 } \over m}
    -{3\over {4m}} q^{2}  }} \sum_{m',m''} D_{m',m} ^{(1/2)*} D_{m''
,m}^{(1/2) }
\cr
&{}&
\left( \langle p q \alpha \vert (1+P) j_{1 } ^{(1)} \vert
\Psi_{^{3}\mbox{He}}   m' \rangle ^{* } 
+ \langle p
  q \alpha \vert (1+P) j_{-1} ^{(1)} \vert \Psi_{^{3}\mbox{He}} m' 
\rangle ^{*} \right)
\langle p q \alpha \vert U_{\rho} m'' \rangle 
  \Bigr]
\label{eq35}
\end{eqnarray} 

\noindent
Now $j_{1} $ ($j_{-1}$) increases (decreases) the magnetic quantum number by 1.
Consequently 

\begin{eqnarray}
R_{TL'} &=& 
- {3 \over \pi } \mbox{Im} \Biggl[ {\sum \hskip -13 pt \int} { 1 \over {E - i \epsilon -
    { p^{2 } \over m }- { 3 \over {4m} }q^{2}  }} 
\cr
&{}& \Bigl( D^{(1/2)} _{{1\over 2},m} D^{(1/2)*}_{-{1\over 2},m} 
\langle p q \alpha {\cal J}
{1 \over 2 } \vert (1+P) \rho ^{(1)} \vert  \Psi
_{^{3}\mbox{He} } {1 \over 2} \rangle 
\langle p q \alpha {\cal J}
 {1 \over 2} \vert U_{j_{1}} -{1\over 2} \rangle ^{*} 
\cr
&+&D^{(1/2)} _{-{1\over 2},m} D^{(1/2)*}_{ {1\over 2},m} 
\langle p q \alpha {\cal J
  } -{1 \over 2 } \vert (1+P) \rho ^{(1)} \vert  \Psi
_{^{3}\mbox{He} } -{1 \over 2} \rangle 
\langle p q \alpha {\cal J} -{1 \over 2} \vert U_{j_{-1}} {1\over 2} \rangle ^{*}
\Bigr)
\cr 
 & - &
 { 1 \over {E + i \epsilon -
    { p^{2 } \over m }- { 3 \over {4m} }q^{2}  }}
\Bigl( D^{(1/2)*} _{-{1\over 2},m} D^{(1/2)}_{{1\over 2},m} 
\langle p q \alpha {\cal J} {1 \over 2 } \vert (1+P) j_{1} ^{(1)} \vert  \Psi
_{^{3}\mbox{He} } -{1 \over 2} \rangle ^{*} 
\langle p q \alpha {\cal J} {1 \over 2} \vert U_{\rho} {1\over 2} \rangle  
\cr
&+&D^{(1/2)*} _{{1\over 2},m} D^{(1/2)}_{-{1\over 2},m} 
\langle p q \alpha {\cal J} -{1 \over 2 } \vert (1+P) j_{-1} ^{(1)} \vert  \Psi
_{^{3}\mbox{He} } {1 \over 2} \rangle  ^{*}
\langle p q \alpha {\cal J} -{1 \over 2} \vert U_{\rho} -{1\over 2} \rangle 
\Bigr)
  \Biggr]
\label{eq36}
\end{eqnarray}

Again we use phase relations
\begin{eqnarray}
\langle p q \alpha {\cal J} -{1 \over 2} \vert ( 1+ P) j_{-1}^{(1)} \vert
\Psi_{^{3}\mbox{He}} {1 \over 2} \rangle = 
(-1)^{{\cal J}
-{1 \over 2}} \Pi \langle p q \alpha {\cal J} {1\over 2} \vert (1 + P ) j_{1}
^{(1)} \vert \Psi _{^{3}\mbox{He}} -{1 \over 2} \rangle 
\label{eq37}
\end{eqnarray}
\begin{eqnarray}
\langle p q \alpha {\cal J} -{1 \over 2} \vert U_{\rho} -{1\over 2} \rangle 
= (-1)^{{\cal J}
-{1\over 2}} \Pi \langle p q \alpha {\cal J}{ 1 \over 2 } \vert U_{\rho}
{1 \over 2} \rangle 
\label{eq38}
\end{eqnarray}
and can simplify $R_{TL'}$ as 

\begin{eqnarray}
R_{TL'} &=& - { 3 \over \pi } \mbox{Im} {\sum \hskip -13 pt \int} \Biggl[ 
{1 \over { E - i \epsilon - { p^{2} \over m} -{ 3 \over {4m}} q^{2} }} 
\langle p q \alpha {\cal J}
 {1 \over 2 } \vert ( 1 + P ) \rho ^{(1)} \vert \Psi_{^{3}
  \mbox{He}} {1\over 2} \rangle 
\cr 
&{}&
\langle p q \alpha {\cal J}
 {1 \over 2} \vert U_{j_{1}} -{1 \over 2} \rangle ^{*} 
\Bigl( D ^{(1/2)} _{{1\over 2},m} D ^{(1/2)*} _{-{1\over 2},m} + 
D^{(1/2)}_{-{1\over 2},m}  D^{(1/2)*}_{ {1\over 2},m } \Bigr)
\cr
&{}& 
-{1 \over { E + i \epsilon - { p^{2} \over m} -{ 3 \over {4m}} q^{2} }} 
\langle p q \alpha {\cal J}
 {1\over 2} \vert ( 1+ P) j_{1}^{(1)} \vert
\Psi_{^{3}\mbox{He}} - {1\over 2 } \rangle ^{*}
\cr
&{}&
\langle p q \alpha {\cal J} {1\over 2} \vert U_{\rho} {1 \over 2} \rangle
\left( D^{(1/2)*} _{-{1\over 2}, m  } D^{(1/2)}_{{1\over 2}, m} + 
D^{(1/2)*}_{{1\over 2},m  } D ^{(1/2) } _{-{1\over 2},m} \right)
\Biggr]
\cr
&=&
-{ 3 \over \pi} \mbox{Im} {\sum \hskip -13 pt \int} \Biggl[
{1 \over { E - i \epsilon - { p^{2} \over
      m} -{ 3 \over {4m}} q^{2} }}
\langle p q \alpha {\cal J} {1 \over 2} \vert (1 + P) \rho ^{(1)} \vert \Psi
_{^{3}\mbox{He}} { 1\over 2} \rangle \langle p q \alpha {\cal J}{ 1 \over 2} \vert 
U_{j_{1}} -{1 \over 2} \rangle ^{*} 
\cr
&-&
{1 \over { E + i \epsilon - { p^{2} \over
      m} -{ 3 \over {4m}} q^{2} }}
\langle p q \alpha {\cal J}{1\over 2} \vert ( 1 + P) j_{1} ^{(1)} \vert \Psi
_{^{3}\mbox{He }} - {1\over 2} \rangle ^{* } 
\langle p q  \alpha {\cal J}{ 1 \over 2} \vert U_{\rho } {1 \over 2} \rangle 
\Biggr]
\cr
&{}&
\left( D^{(1/2)} _{{1 \over 2} ,m} D^{(1/2)*} _{-{1\over 2},m  } 
+ D ^{(1/2)}_{-{1\over 2},m } D^{(1/2)*}_{{1 \over 2},m} \right)
\cr
&=&
\pm { 3 \over \pi } \mbox{Im} {\sum \hskip -13 pt \int} [ { 1 \over { E+ i \epsilon -{ p^{2}
      \over m } -{ 3 \over {4m} } q^{2} }} 
\langle p q \alpha {\cal J}{1 \over 2 } \vert ( 1+ P) j_{1 } ^{(1)} \vert \Psi
_{^{3}\mbox{He}} -{1 \over 2} \rangle ^{*} \langle p q \alpha {\cal J}
{1 \over 2}
\vert U_{\rho } { 1 \over 2} \rangle 
\cr 
&-&
{ 1 \over { E - i \epsilon -{ p^{2 }\over m} -{ 3 \over {4m} }q^{2} } } 
\langle p q \alpha {\cal J}{1\over 2} \vert (1+ P) \rho  ^{(1)} \vert \Psi
_{^{3}\mbox{He}} {1 \over 2} \rangle  \langle p q \alpha {\cal J}{ 1 \over 2 }
\vert U_{j_{1} }  -{1 \over 2} \rangle ^{*} ] 
\sin{\theta ^{*}} \cos{\phi ^{*}}
\label{eq39}
\end{eqnarray}
where $\pm$ refers to $m=\pm {1\over 2} $, respectively.
As shown in the Appendix one gets similarly
\begin{eqnarray}
R_{T'}&=& \pm \cos{\theta ^{*} } {3 \over \pi } \mbox{Im} {\sum \hskip -13 pt \int} \Biggl[ 
{ 1 \over { E + i \epsilon -{ p^{2} \over m} - {3 \over {4 m} } q^{2} }}
\cr 
& {} &
\Bigl( \langle p q \alpha {\cal J} {1\over 2 } \vert (1+P) j_{1}^{(1)} \vert
  \Psi_{^{3}\mbox{He}} -{1\over 2} \rangle ^{*} 
\langle p q \alpha {\cal J}
{ 1 \over 2 } \vert U_{j_{1}}  -{1\over 2} \rangle
\cr 
&-& \langle p q \alpha {\cal J} { 3 \over 2} \vert (1 + P) j_{1 } ^{(1)} \vert 
\Psi_{^{3}\mbox{He} } {1 \over 2} \rangle ^{*} \,  \langle p q \alpha { \cal J} { 3
  \over 2} \vert U_{j_{1}} {1 \over 2} \rangle  \Bigr)\Biggr]
\label{eq40}
\end{eqnarray}
The partial wave projected matrix elements are evaluated according to our
standard techniques \cite{cite12,cite13,cite14,cite15,cite16}.

The only structure functions depending on $\theta ^{*} $ and $\phi ^{*}$ are 
\begin{eqnarray}
R_{T'} \equiv \tilde R_{T'} \cos{\theta ^{*}}
\label{eq41}
\end{eqnarray}
\begin{eqnarray}
R_{{TL'}} \equiv \tilde R_{{TL'}} \sin{\theta ^{*}} \cos{\phi ^{*}}
\label{eq42}
\end{eqnarray}
Then according to Eqs.~(\ref{eq1}) and (\ref{eq3}) the asymmetries are 
\[
A \equiv
{
{
  {{d \sigma } \over {d \hat k ' d k'_{0}} } \Bigl \vert _{h=1}  
 -{{d \sigma } \over { d \hat k' d k'_{0}} } \Bigl \vert _{h=-1} 
}
\over 
{
  {{d \sigma } \over {d \hat k ' d k'_{0}} } \Bigl \vert _{h=1}  
 +{{d \sigma } \over { d \hat k' d k'_{0}} } \Bigl \vert _{h=-1} 
}
}
\]
\begin{eqnarray}
= { {v_{T'} \tilde R_{T'} \cos{\theta ^{*}} + v_{TL'} \tilde R_{TL'}
    \sin{\theta ^{*}} \cos{\phi ^{*}}  } \over { v_{L} R_{L} + v_{T} R_{T}}}
\label{eq43}
\end{eqnarray}

Putting the angle $\theta ^{*}$ between the direction of the $^{3}$He target
spin ($m={1\over 2}$) and the direction $\hat Q$ of the virtual photon to
zero one selects the transverse asymmetry $A_{T'}$ (proportional  to $\tilde
R_{T'}$), whereas putting that angle to 90$^{\circ}$ one gets the
transverse-longitudinal asymmetry $A_{TL'}$ (proportional to $\tilde R_{TL'}$).

Let us now regard the most simplified picture. We neglect all final state 
interactions,
thereby excluding also the pd break up channel. Also the antisymmetrization 
is kept only in the two-body subsystem described by $ \vec p $. Finally
we restrict the $^3$He wave function
to the principal S-state. 
In order to define clearly our notation we start from the matrix elements for 
the symmetrized plane wave impulse approximation PWIAS 
\begin{eqnarray}
N^\mu _{PWIAS} &\equiv& { 1\over \sqrt{3 !}} \langle \vec p \vec q m_1 m_2 m_3 
\tau_1 \tau_2 \tau_3 \vert (1 - P_{23} ) ( 1 + P ) j^{\mu}( \vec Q ) 
\vert \Psi _{^{3}\mbox{He}} m \rangle _{ \theta ^* \phi ^*} 
\cr
&=&
{ 3 \over \sqrt{3!}} \langle \vec p \vec q m_1 m_2 m_3 \tau_1 \tau_2 \tau_3 
\vert (1 - P_{23}) ( 1 + P ) j^\mu _{(1)} (\vec Q)\vert \Psi _{^{3}\mbox{He}} m 
\rangle  _{ \theta ^* \phi ^ * }
\label{extra}
\end{eqnarray}
As before we reduced the single nucleon current operator to one term. 
The subscript $(1)$ indicates the particle number, which in our notation 
is described by $\vec q$. Now we drop the permutation operator $P$, 
apply $P_{23}$ and insert the principal S-state approximation.
The resulting nuclear matrix elements are
          
\begin{eqnarray}
\tilde N_{0 } \equiv \sqrt{6} \langle \vec p \vec q 
m_{1} m_{2} m_{3} 
\tau_{1} \tau_{2} \tau_{3} 
\vert 
{\rho}^{(1)}
(\vec Q) \vert \Psi_{^{3}\mbox{He}} ^{PS}m \rangle _{\theta ^{*} \phi ^{*}} 
\label{eq44}
\end{eqnarray}
\begin{eqnarray}
\tilde N_{\pm} \equiv \sqrt{6} \langle\vec p \vec q 
m_{1} m_{2} m_{3} 
\tau_{1} \tau_{2} \tau_{3} 
\vert 
j_{\pm}^{(1)}
(\vec Q) \vert \Psi_{^{3}\mbox{He}} ^{PS} m \rangle _{\theta ^{*} \phi ^{*}} {\rm ,}
\label{eq45}
\end{eqnarray}

The principal S-state is
           
\begin{eqnarray}
\vert \Psi_{^{3}\mbox{He}} ^{PS} m \rangle  = \vert \phi _{S} \rangle 
\vert \xi_{a} m
\rangle {\rm ,}
\label{eq46}
\end{eqnarray}
where
$\vert \xi_{a } m \rangle $ is the totally antisymmetrical spin-isospin state
\begin{eqnarray}
\vert \xi _{a} m \rangle = 
{ 1 \over \sqrt{2}} \bigl( \vert ( t=0 {1 \over 2})
T={1 \over 2} \rangle \vert (s=1 {1 \over 2}) S={1\over 2} m \rangle 
- \vert (t=1 {1\over 2}) T={1\over 2 } \rangle \vert (s=0 {1\over 2} )
S={1\over 2 } m \rangle \bigr)
\label{eq47}
\end{eqnarray}
and $\vert \phi _{S} \rangle $ is the totally symmetrical space part belonging
to total orbital angular momentum $L=0$.
In terms of our standard notation~\cite{cite18} one easily gets
\[
\vert \Psi_{^{3}\mbox{He}}^{PS} m \rangle  = 
\]
\begin{eqnarray}
\sum _{l \ {\rm even}} \sum_{s, t }
\int dp p^{2} \int dq q^{2} 
\vert p q (l l )0 (s {1 \over 2} ) S={1\over 2} m (t {1 \over 2}) T={1 \over 2}
\rangle \phi _{l} (p q) {1 \over \sqrt{2}} \Bigl( \delta_{s 1} \delta_{ t 0} -
\delta _{ s 0} \delta _{t 1} \Bigr) 
\label{eq48}
\end{eqnarray}
with 
\begin{eqnarray}
\phi_{l} (p q ) = 
{ 1 \over \sqrt{2}} \Biggl( 
\Psi _{ (l l) 0 (1 { 1\over 2} ) {1\over 2}  (0 {1 \over 2} )
  {1\over 2}} (p q ) - 
\Psi _{(l l ) 0  (0 {1 \over 2}) {1\over 2} 
(1 { 1 \over 2}) { 1 \over 2} } (p q ) 
\Biggr)
\label{eq49}
\end{eqnarray} 
and $\Psi _{\alpha } (p q )$ are the wave function components $\langle p q
\alpha \vert \Psi m \rangle $ determined in the Faddeev scheme.
Using (\ref{eq24}) and (\ref{eq46}) the nuclear matrix elements 
(\ref{eq44}) and  (\ref{eq45}) turn into 

\begin{eqnarray}
{\tilde{N}}_{0} = 
\sqrt{6} 
F_1^{( \tau_1 ) } ( \vec Q ) 
\sum_{m'} 
D ^{(1/2)} _{m' , m}   \ 
\phi_{S} ( \vec p , \vec q - \frac23 \vec Q ) \
\langle  m_1 m_2 m_3 \tau_1 \tau_2 \tau_3 \vert  \xi_{a } m' \rangle 
\label{eq50}
\end{eqnarray}

\[
{\tilde{N}}_{\pm 1} = 
\sqrt{6} 
F_1^{( \tau_1 ) } ( \vec Q ) \,
\frac{ q_{\pm 1}  }{m_N} \,
\sum_{m'} 
D ^{(1/2)} _{m' , m}   \ 
\phi_{S} ( \vec p , \vec q - \frac23 \vec Q ) \
\langle  m_1 m_2 m_3 \tau_1 \tau_2 \tau_3 \vert  \xi_{a } m' \rangle 
\]
\begin{eqnarray}
- \
\sqrt{12} G_M^{( \tau_1 ) } ( \vec Q ) \,
\frac{ \vert \vec Q \vert }{2 m_N} 
\sum_{m'} 
D ^{(1/2)} _{m' , m}   \ 
\phi_{S} ( \vec p , \vec q - \frac23 \vec Q ) \
\langle  m_1 \mp 1 \, m_2 m_3 \tau_1 \tau_2 \tau_3 \vert  \xi_{a } m' \rangle 
\label{eq51}
\end{eqnarray} 

Thereby the single particle current operator has been chosen 
according to \cite{cite13}.
Despite the approximate, not fully antisymmetrized final state in (\ref{eq44}) and 
(\ref{eq45}), we stick to the summation prescription over all final states
in the evaluation of the structure functions, which corresponds to the fully
antisymmetrized final states in Eq.~(\ref{extra}):
\begin{eqnarray}
{ \sum \hskip -13 pt \int } 
\equiv { 1 \over 6} \sum _{ m_1 m_2 m_3 } \sum
 _{\tau_1 \tau_2 \tau_3} \int d \vec p  d \vec q 
\end{eqnarray}
Then a straightforward evaluation yields

\begin{eqnarray}
R_L = \frac{2 m_N}{3}  \int_0^{p_{\rm max}} d p p^2 q \int d \hat p \, \int d \hat q  \,
{\vert  \phi_{S} ( \vec p , \vec q - \frac23 \vec Q ) \vert}^2 \, 
\left( \frac13  ( F_1^{(n)} ( \vec Q ))^2 + 
 \frac23  ( F_1^{(p)} ( \vec Q ))^2 \right)
\label{eq52}
\end{eqnarray} 

\[
R_T = \frac{2 m_N}{3}  \int_0^{p_{\rm max}} d p p^2 q \int d \hat p \, \int d \hat q  \,
{\vert  \phi_{S} ( \vec p , \vec q - \frac23 \vec Q ) \vert}^2 \, 
\]
\begin{eqnarray}
\left[  \frac{8 \pi }{9} 
\frac{ {\vert \vec q \vert }^2 }{m_N^2} \,
{\vert Y_{ 1 , 1 } ( \hat q ) \vert}^2 \, 
 \left( ( F_1^{(n)} ( \vec Q ))^2 + 2 ( F_1^{(p)} ( \vec Q ))^2 \right) \,
+
 \left( \frac23 ( G_M^{(n)} ( \vec Q ))^2 + \frac43 ( G_M^{(p)} ( \vec Q ))^2 \right) 
\frac{ {\vert \vec Q \vert }^2 }{4 m_N^2} \right]
\label{eq53}
\end{eqnarray} 

\begin{eqnarray}
R_{T'} = \frac{2 m_N}{3}  \int_0^{p_{\rm max}} d p p^2 q \int d \hat p \, \int d \hat q  \,
{\vert  \phi_{S} ( \vec p , \vec q - \frac23 \vec Q ) \vert}^2 \, 
(-\frac16 \cos \theta ^{*} ) 
( G_M^{(n)} ( \vec Q ) )^2
\frac{ {\vert \vec Q \vert }^2 }{m_N^2}
\label{eq54}
\end{eqnarray} 

\begin{eqnarray}
R_{TL'} = \frac{2 m_N}{3}  \int_0^{p_{\rm max}} d p p^2 q \int d \hat p \, \int d \hat q  \,
{\vert  \phi_{S} ( \vec p , \vec q - \frac23 \vec Q ) \vert}^2 \, 
\frac{\sqrt{2}}{3}  F_1^{(n)} ( \vec Q ) \,  G_M^{(n)} ( \vec Q )
\frac{ {\vert \vec Q \vert } }{m_N}
\cos \phi ^{*} \, \sin \theta ^{*} 
\label{eq55}
\end{eqnarray} 

The energy conserving delta function gives $p_{\rm max} $ and $q$ to be
\begin{eqnarray}
p_{\rm max} = \sqrt { m_N \, E }
\label{eq56}
\end{eqnarray} 

\begin{eqnarray}
q \, = \, \sqrt { \frac43 ( p_{\rm max}^2 - p^2 ) } {\rm .}
\label{eq57}
\end{eqnarray} 
Note that $R_L$ and $R_T$ receive contributions from neutrons 
and protons, whereas due to the principal S-state  assumption
$ R_{T'} $ and $ R_{TL'}$ are fed only by the neutron contribution.
It results in the asymmetry 

\[
A \, = \, 
\left[ 
v_{T'} \, (-\frac16 \cos \theta ^{*} ) \,
( G_M^{(n)} ( \vec Q ) )^2
\frac{ {\vert \vec Q \vert }^2 }{m_N^2} \ + \
v_{TL'} \, 
\frac{\sqrt{2}}{3}  
F_1^{(n)} ( \vec Q ) \,  G_M^{(n)} ( \vec Q ) \,
\frac{ {\vert \vec Q \vert } }{m_N} \,
\cos \phi ^{*} \, \sin \theta ^{*} \right] 
\]
\[
\left.  \Biggl/ \ \ \    \Biggl[
v_L \left( \frac13( F_1^{(n)} ( \vec Q ))^2 + \frac23 ( F_1^{(p)} ( \vec Q ))^2 \right) \,
+ \right.
\]
\begin{eqnarray}
\left.
v_T \, \left\{
\left( ( F_1^{(n)} ( \vec Q ))^2 + 2 ( F_1^{(p)} ( \vec Q ))^2 \right) 
 \alpha ( \omega , \vert \vec Q \vert )
+
\left( \frac23 ( G_M^{(n)} ( \vec Q ))^2 + \frac43 ( G_M^{(p)} ( \vec Q ))^2 \right) 
\frac{ {\vert \vec Q \vert }^2 }{4 m_N^2}
\right\}
\right]
{\rm ,}
\label{eq58}
\end{eqnarray} 

\noindent
where 

\[
\alpha ( \omega , \vert \vec Q \vert ) \, = \,
\frac{ \frac{8 \pi}{9} \,
\int_0^{p_{\rm max}} d p p^2 q \int d \hat p \, \int d \hat q  \,
{\vert  \phi_{S} ( \vec p , \vec q - \frac23 \vec Q ) \vert}^2 \, 
\frac{ {\vert \vec q \vert }^2 }{m_N^2}
{\vert Y_{ 1 , 1 } ( \hat q ) \vert}^2 
}
{
\int_0^{p_{\rm max}} d p p^2 q \int d \hat p \, \int d \hat q  \,
{\vert  \phi_{S} ( \vec p , \vec q - \frac23 \vec Q ) \vert}^2 \, 
}
\]
\begin{eqnarray}
= \ \frac13 \,
\frac{
\int_0^{p_{\rm max}} d p p^2 q 
\frac{q^2}{m_N^2} \, \sum_l \,
\int_{-1}^1  d x \, ( 1 - x^2 )  \,
\phi_{l}^2 ( p ,\vert  \vec q - \frac23 \vec Q \vert ) \, 
}
{
\int_0^{p_{\rm max}} d p p^2 q \,  \sum_l \, \int_{-1}^1  d x \, 
\phi_{l}^2 ( p ,\vert  \vec q - \frac23 \vec Q \vert ) 
}
\label{eq59}
\end{eqnarray} 
with $x=\hat q \cdot \hat Q$.

That factor $\alpha ( \omega , \vert \vec Q \vert ) $ is due to the convection current,
whose contribution survives solely in $R_T$ and prevents that the dependence
on the $^3$He wave function drops out. It is typically of the order 10$^{-3}$ 
and together with $ F_1^2 ( \vec Q ) $ of neutron and proton
it is negligible in relation 
to the other term at the momentum transfer $ \vert \vec Q \vert$ considered. 

If we insert the explicit expressions for the kinematical factors $v$ and
use the nonrelativistic approximation $ Q^2 \approx -{\vec Q}^{\ 2} $ we get

\begin{eqnarray}
A \ = \ 
\frac{
\frac{Q^2}{2 m_N^2} \, \tan {\Theta \over 2} \left[
\sqrt{{-Q^{2} \over {\vec Q ^{2}} } + \tan^{2} {\Theta \over 2}}
( G_M^{(n)} )^2 \, \cos \theta ^{*} \ + \
\frac{2 m_N}{\vert \vec Q \vert} \,
F_1^{(n)} 
G_M^{(n)} \,
\cos \phi ^{*} \sin \theta ^{*} \right]
}
{
( F_1^{(n)} )^2
 + 2 ( F_1^{(p)} )^2
 - \frac{Q^2}{4 m_N^2}  
\left[
( G_M^{(n)} )^2 + 2 ( G_M^{(p)} )^2 + 
{\alpha}
\frac{6 m_N^2}{{\vert \vec Q \vert}^2 } 
\left( ( F_1^{(n)} )^2 + 2 ( F_1^{(p)} )^2 \right)
\right] ( 1 + 2 \tan^{2} {\Theta \over 2} )
}
\label{eq60}
\end{eqnarray} 
where we kept $({{-Q^{2} \over {\vec Q ^{2}}}})$ under the square root in
order to facilitate the comparison to the asymmetry gained by scattering
a polarized electron on a polarized nucleon target. That well known expression
is 

\begin{eqnarray}
A_{\rm nuc} = 
\frac{
\frac{Q^2}{2 m_N^2} \, \tan {\Theta \over 2} \left(
\sqrt{{-Q^{2} \over {\vec Q ^{2}} } + \tan^{2} {\Theta \over 2}}
G_M^2 \, \cos \theta ^{*} \ + \
\frac{2 m_N}{\vert \vec Q \vert } \, G_E G_M \,
\cos \phi ^{*} \sin \theta ^{*} \right) ( 1 - \frac{Q^2}{4 m_N^2} )
}
{
G_E^2 - \frac{Q^2}{4 m_N^2}  G_M^2
( 1 + 2  ( 1 - \frac{Q^2}{4 m_N^2} )  )  \tan^{2} {\Theta \over 2} 
}
\label{eq61}
\end{eqnarray} 

The numerators in (\ref{eq60}) and  (\ref{eq61}) are equal 
except that we use $F_1$ instead of $G_E$.
Our single nucleon current operator \cite{cite13} contains $F_1$.
In the denominator of (\ref{eq60}), however, 
there are also contributions from the protons in $^3$He
and the correction term $\alpha$ resulting from the convection 
current. In $^3$He the nucleons are moving in contrast to the case 
of a fixed single nucleon target.

Regarding the expression (\ref{eq60}) we see that the transverse asymmetry 
$A_{T'}$ defined for $\theta^* = 0^o $ is proportional to $( G_M^{(n)} )^2$,
whereas the transverse-longitudinal asymmetry $A_{TL'}$ defined 
for $\theta^* = 90^o $ is proportional to $ F_1^{(n)} \, G_M^{(n)} $.
Will that simple result survive under more realistic conditions ?

This is just the aim of our study to learn how a more realistic 
$^3$He wave function, the inclusion of antisymmetrization in the final 
state and the inclusion of final state interactions among the three
final nucleons modifies that simple picture and whether these
modifications will still leave sufficient sensitivity to the 
value of the magnetic form factor$G_M^{(n)}$ of the neutron.

Let us now define the various levels of evaluating the two asymmetries 
$A_{T'}$ and $A_{TL'}$  .
The form (\ref{eq60}) based on the principal S-state and plane wave 
impulse approximation without antisymmetrization 
in the final state 
(see Eqs. (\ref{eq44})--(\ref{eq45})) will be denoted by PWIA (PS).
If we include the realistic $^3$He wave function we denote the result
by PWIA. The corresponding structure functions are determined 
by (\ref{eq22}) dropping the factor $3$, the permutation operator $P$ 
and $U_B$ and $U_A$ should be chosen by (\ref{eq19}) without the two terms 
proportional to the NN t-matrix $t$. 
If one restricts $ \vert \Psi _{^{3}\mbox{He}} \rangle $ to the principal
S-state  the results should be identical to the structure functions 
evaluated
according to 
Eqs.~(\ref{eq52})--(\ref{eq55})
and to the asymmetry from (\ref{eq60}). This is a very nontrivial check 
and turned out to be very well fulfilled. 

The next improvement of the theory is to keep plane waves
in the final state but antisymmetrize them correctly. This is achieved
using  (\ref{eq22}) and dropping only in 
the $U$-amplitudes of Eq.~(\ref{eq19})
the terms proportional to $t$. This approximation will be denoted
by PWIAS. 

An intermediate step for including the full final state interaction 
is to keep in the nuclear matrix elements the interaction in the 
pair of nucleons which are spectators to the absorption process of the photon
on the third nucleon. This approximation is described by the nuclear 
matrix elements 

\begin{eqnarray}
{N_{0 }}'  = \sqrt{6} \, \langle \vec p \vec q m_{1}' m_{2}' m_{3}' \vert 
 ( 1 + t G_0 ) 
\rho^{(1)}
(\vec Q) \vert \Psi_{^{3}\mbox{He}} m \rangle _{\theta ^{*} \phi ^{*}} 
\label{eq62}
\end{eqnarray}
\begin{eqnarray}
{N_{\pm}}' = \sqrt{6} \, \langle\vec p \vec q m_{1}' m_{2}' m_{3}' \vert 
 ( 1 + t G_0 ) 
j _{\pm}^{(1)}
(\vec Q) \vert \Psi_{^{3}\mbox{He}} m \rangle _{\theta ^{*} \phi ^{*}} 
\label{eq63}
\end{eqnarray}

\noindent
for the ppn-breakup process and by 

\begin{eqnarray}
{N_{0, d }}'  = \langle \varphi_d  \vec q m_{1}' m_{d}'  \vert 
\rho^{(1)}
(\vec Q) \vert \Psi_{^{3}\mbox{He}} m \rangle _{\theta ^{*} \phi ^{*}}
\label{eq64}
\end{eqnarray}
\begin{eqnarray}
{N_{\pm 1, d }}' =  \langle \varphi_d \vec q m_{1}' m_{d}' \vert              
j _{\pm}^{(1)}
(\vec Q) \vert \Psi_{^{3}\mbox{He}} m \rangle _{\theta ^{*} \phi ^{*}}
\label{eq65}
\end{eqnarray}

\noindent
for the pd-breakup process. Note that we did not antisymmetrize 
the final state except in the two-body subsystem.
This leads to the expression (\ref{eq22}) without the factor $3$ and the permutation 
operator $P$, and the $U$-amplitudes are just given by the driving term 
in  Eq.~(\ref{eq19}). The corresponding results will be denoted by PWIA'.
If on top of that we antisymmetrize the final state the result will 
be denoted by PWIAS'. This is evaluated using Eq.~(\ref{eq22})
as it is, but the $U$-amplitude as for  PWIA'.

Finally evaluating (\ref{eq22}) and (\ref{eq19}) exactly and thus 
including the final state interaction to all orders and between all
three nucleons, as well as including the antisymmetrization fully
will be denoted by FULL.

\section{Results}

We used the Bonn B NN potential~\cite{cite19}
and kept its force components up to
total two-nucleon angular momentum j=2 in the treatment of the 3N continuum.
The effects of the j=3 components stayed below the percentage level.
The electromagnetic nucleon form factors
are from \cite{cite20}.

The experimental setup for the spin-dependent asymmetry can be 
characterized by 
the initial electron energy ($k_0$), 
the electron scattering angle ($\Theta$),
two angles which parametrize the direction of the target 
polarization ($\theta_A$, $\phi_A$) 
(see Fig. 7 of Ref. \cite{cite10}, e.g.), and the measured energy 
transfer ($\omega$).
These values used in the recent experiments 
\cite{cite7,cite8,cite9} are summarized in Table \ref{table1}, 
together with energy transfer ($\omega_{QE}$), 3-momentum transfer 
(${\vert \vec Q \vert }_{QE}$) and the angles defining 
the polarization with respect 
to the direction $ \hat Q$ of the 3-momentum transfer ($\theta^*_{QE}$ and 
$\phi^*_{QE}$) at the quasielastic (QE) condition.
The asymmetry measured in Ref. \cite{cite7} near the quasielastic 
kinematics is essentially the transverse asymmetry $A_{T^\prime}$ 
because of the condition, $\theta^* \simeq$ 0\cdeg, 
and then is expected to be sensitive to the neutron magnetic form 
factor.
Thus hereafter the asymmetry measured in this experiment will be 
referred to as simply $A_{T^\prime}$.
On the other hand, those measured near the quasielastic 
kinematics \cite{cite8} and a lower-$\omega$ region just above 
the 3-body breakup threshold \cite{cite9} are essentially 
the transverse-longitudinal asymmetry $A_{TL^\prime}$
because of the condition, $\theta^* \simeq$ 90\cdeg, 
and then are expected to be sensitive to both of the neutron charge 
and magnetic form factors.
Hereafter  the asymmetry measured in these experiments will be 
referred to as simply $A_{TL^\prime}$.
These experimental results were analyzed by recent theoretical 
works \cite{cite10,cite11} with realistic \hel{} wave functions and 
plane wave impulse approximation. 
In this article we call that approximation PWIA'.
In Ref. \cite{cite7}, the neutron magnetic form factor, $G_M^n$, 
was extracted based on PWIA' with reasonable 
agreement with experimental data.
On the other hand, agreement between the PWIA' calculations and 
the measured asymmetries in Refs. \cite{cite8,cite9} is rather poor.
The PWIA' prediction of the asymmetry in the quasielastic region was 
found to be large compared to the experimental data \cite{cite8} 
at the $(1-2.5) \sigma$ level.
At the lower $\omega$-region \cite{cite9}, the 
experimental asymmetry was found 
to be enhanced in contradiction with PWIA' calculations.

\begin{figure}
\input{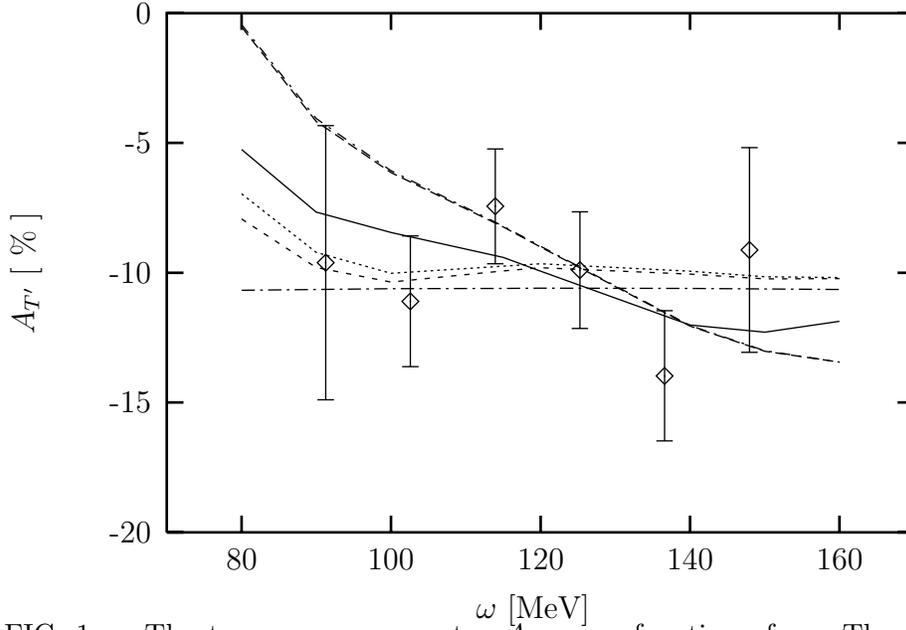}

\bigskip
\bigskip

\caption{
The transverse asymmetry $A_{T'}$ as a function of $\omega$. The data are from
Ref\protect\cite{cite7}. The six theoretical curves are
PWIA(PS)(dashed-dotted),  
PWIA(dotted), PWIAS(short dashed), PWIA'(long dashed),
PWIAS'(dashed-dotted, declined curve) and FULL (solid). Note PWIA' and PWIAS'
overlap.
}
\label{fig1}
\end{figure}

\begin{figure}
\input{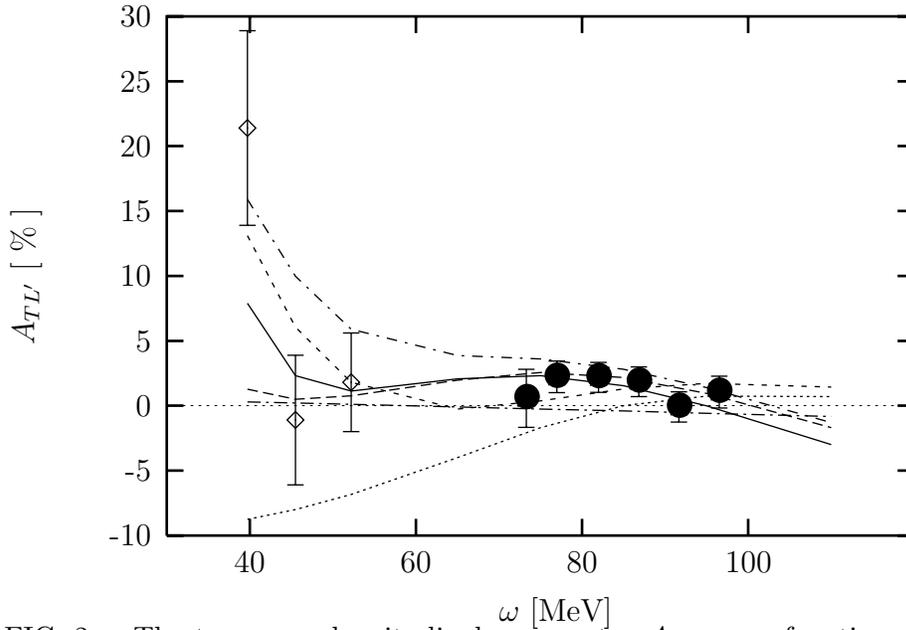}

\bigskip
\bigskip

\caption{
The transverse-longitudinal asymmetry $A_{TL'}$ as a function of $\omega$. The
data ($\diamond$) are from Ref\protect\cite{cite8} and the data
 ({\Huge $\bullet$}) from
Ref\protect\cite{cite9}. Curves as in Fig 1.
The PWIAS'-curve rises to the data point at $\omega$=40MeV.
}
\label{fig2}
\end{figure}

Let us now regard our results in comparison to the experimental data
for $A_{ T'}$
in Fig.~\ref{fig1} and for $A_{ TL'} $
in Fig.~\ref{fig2}. We display six theoretical curves.
The most naive prediction, PWIA(PS) 
lies within the error bars for four of the six data points for $A_{T'}$
in Fig.~\ref{fig1}. In case of  $A_{TL'}$  shown in Fig.~\ref{fig2}
that prediction is essentially zero and clearly disagrees with the data.
Replacing the principal S-state approximation of $^3$He by the full
expression,
called PWIA, causes a visible change 
for  $A_{T'}$ at low $\omega $'s and a much larger one for  $A_{TL'}$.
Now for  $A_{TL'}$ one deviates even stronger from the data.
Apparently  $R_{TL'}$ is more sensitive to the $^3$He wave function
than  $R_{T'}$. Symmetrizing the final state using PWIAS has a small
effect for  $A_{T'}$ but a big one on  $A_{TL'}$. It rises  $A_{TL'}$
for small $\omega $'s qualitatively similar to what happens 
in the data but misses the data around $\omega = $ 60--70 MeV.
A strong move occurs by keeping the final state interaction among 
the two spectator nucleons, PWIA'. For lower $\omega$'s it appears 
to be somewhat to high for $A_{T'}$ and again at low $\omega$'s 
near the threshold for 3N breakup it does not show the quick rise 
of the one data point in $A_{TL'}$. 
However between 50 and 100~MeV it follows the data  for  $A_{TL'}$.
Now symmetrizing in addition the final state, PWIAS', it does not cause
a visible change  for $A_{T'}$, but overshoots now the data for $A_{TL'}$
for $\omega $'s below about 70~MeV.
Finally the full calculation leads again to a strong shift and agrees  now 
quite well with the data for both $A_{T'}$ and $A_{TL'}$. 
At very low  $\omega $'s it now follows the experimental trend 
for $A_{TL'}$ though still misses the error bar of the last data point 
to the left.
More precise data for $A_{TL'}$, especially in that region 
would be of interest to quantitatively challenge our present day 
understanding of final state interactions but possibly also effects related 
to the choice of the current operator.

Though the data show still some scatter for $A_{T'}$
 we would like to quantify 
these results by providing a $\chi ^{2} $ for  $A_{T'}$:

\begin{equation}
\chi^{2} \, \equiv \,
\sum\limits_i 
\frac{
\left( A_{T'}^{\rm theory} (i) -  A_{T'}^{\rm exp} (i) \right) ^2 
}
{
\left(  A_{T'}^{\rm exp} (i) \right) ^2 
}
\label{eq66}
\end{equation}

The sum runs over the six data points.
They are 4.2, 4.1, 4.0, 6.1, 6.3, 3.4 for PWIA(PS), PWIA, PWIAS, PWIA', 
PWIAS' and FULL, respectively. The FULL calculation describes the data 
best and the correct antisymmetrization and the treatment of the full 
final state interaction is required to achieve quantitative insight.
Note that the often used plane wave impulse approximation, here called 
PWIA' is insufficient.

The aim of the experiments were to achieve information on the magnetic
neutron form factor. 
Therefore the influence of the badly known electric form factor 
of the neutron, $G_E^{(n)}$, or in our nonrelativistic form  $F_1^{(n)}$
should be known. 
We restrict our investigation  to $A_{T'}$ and in addition to 
PWIA(PS) and PWIAS. 
As an extreme assumption we put $F_1^{(n)} $  to zero,
the effect on  $A_{T'}$  was negligible (below 1~\%).
We expect that this remains true even for the FULL calculation and 
therefore we expect that the specific choice of $F_1^{(n)}$ 
will not influence significantly the extraction of information
on $G_M^{(n)}$ from $A_{T'}$.

We add the remark that this extreme assumption puts 
$A_{TL'} = 0 $ for PWIA(PS), of course.
Obviously the data are different from zero and  $A_{TL'}$
receives contributions from ingredients, which go beyond that most
simplistic picture. 
This can already be seen comparing PWIA(PS) and PWIA in Fig.~\ref{fig2}.
The difference is just the replacement of the principal S-state 
$^3$He wave function by 
the realistic one.
Apparently the S'- and D-state pieces contribute very strongly 
to $A_{TL'}$. This was noticed before in~\cite{cite10}.

\begin{figure}
\input{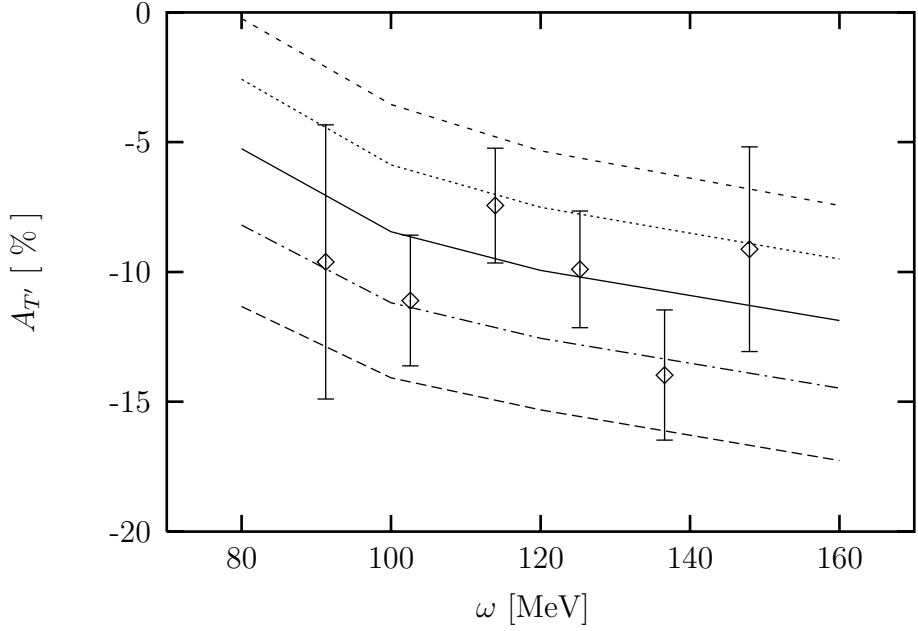}

\bigskip
\bigskip

\caption{
The dependence of the transverse asymmetry $A_{T'} $ in the FULL calculation
on the strength factor $f$ multiplied to the neutron magnetic form factor
$G_{M}^{(n)}$ from \protect\cite{cite20}. 
$f$=0.7 (short-dashed), $f$=0.85 (dotted), 
$f$=1 (solid), $f$=1.15 (dashed-dotted) and $f$=1.3 (long-dashed).
Comparison to data from \protect\cite{cite7}. 
}
\label{fig3}
\end{figure}

\begin{figure}
\input{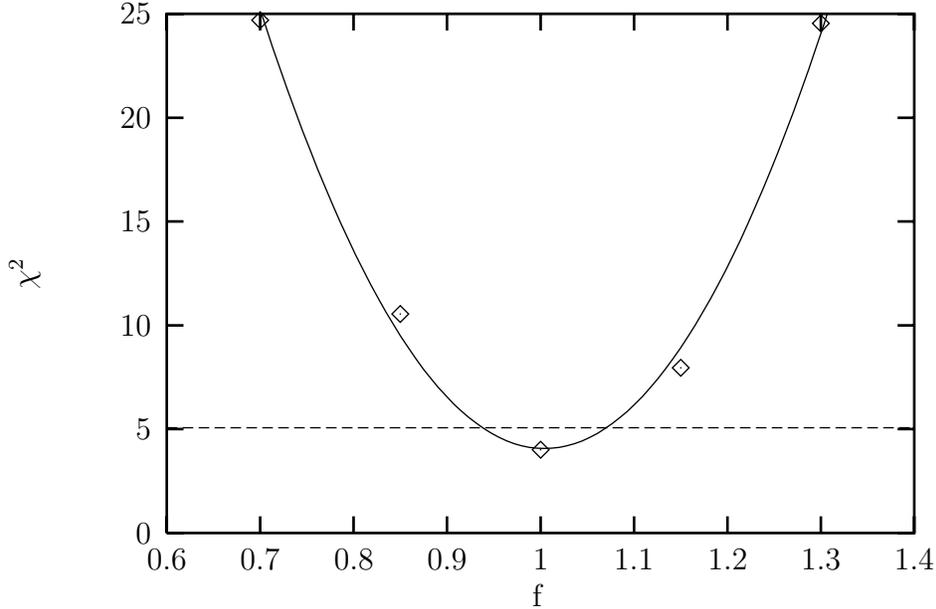}

\bigskip
\bigskip

\caption{
The $\chi^{2}$ from Eq(\protect\ref{eq66}) for $A_{T'}$ as a function of the strength
factor $f$ from Fig. 3. A parabola is fitted to the calculated values denoted
by ($\diamond$). 
The value $\chi ^2_{min}$ +1 is shown as dashed horizontal line and provides 
a spread of $\Delta f =\pm$6.6\%.
}
\label{fig4}
\end{figure}

Being free of that dependence on $F_1^{(n)}$ for $A_{T'}$,
  we now altered 
the neutron magnetic form factor by $\pm $ 15 \%{} and $\pm $ 30 \%{}
and achieved the results,  for the FULL calculation
displayed in  Fig.~\ref{fig3}.
Clearly  $\pm $ 30 \%{} changes lie outside the bulk of the data and also 
$\pm $ 15 \%{}  changes are not acceptable given the data.
One can quantify these studies and extract the optimal $f$ factor
multiplying the neutron magnetic form factor $G_M^{(n)}$ of \cite{cite20}
such that $\chi^{2} $ is minimal. 
This study was performed for the FULL calculation. We display the resulting 
$\chi ^{2} $ in  Fig.~\ref{fig4} 
and extract the optimal $f$ factor to be 1.
As a measure of the accuracy of extracting that value we take 
the spread in $f$ for $\chi^{2}_{\rm min} + 1$. This is $\pm $ 6.6 {\%}.
Clearly more precise data would be very welcome to improve on the accuracy 
of extracting information on  $G_M^{(n)}$.

The possibly most serious theoretical uncertainty
in our analysis is that we do not take
MEC s into account. Their quantitative contribution
remains to be investigated.
It also remains to be seen whether different choices of NN forces could 
change the results. For inclusive scattering without polarisation 
we found only a very  weak dependence \cite{cite16}.
Simplified calculations keeping only $j_{\rm max} = 1 $ NN force components,
now for the polarization case, also did not show a dependence 
on the choice of the NN force.

\begin{figure}
\input{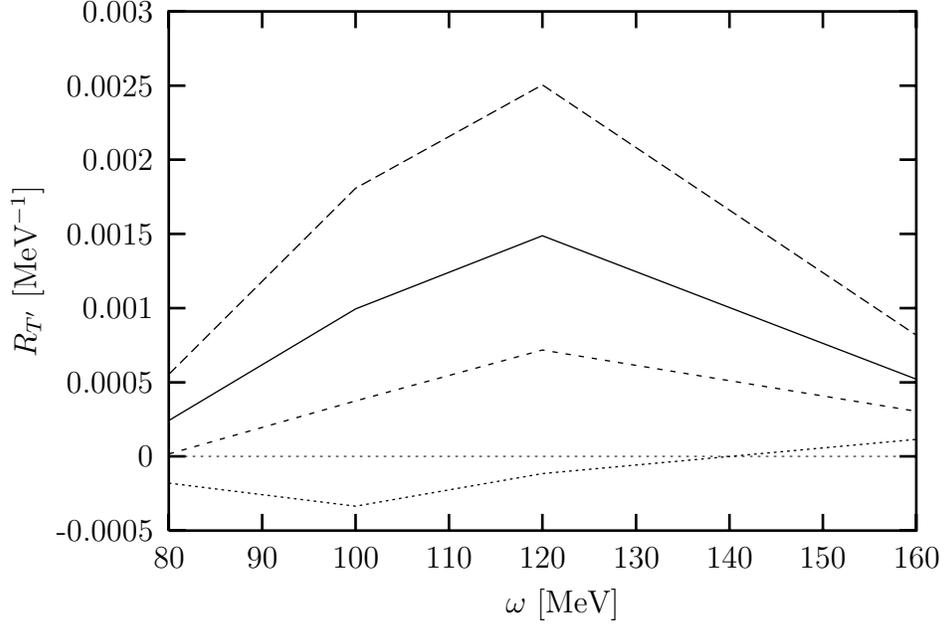}

\bigskip
\bigskip

\caption{
The transversal structure function $R_{T'}$ as a function of $\omega$ in the
FULL calculation for various strength factors $f$: $f$=1.3 (long dashed), 
$f$=1 (solid),$f$=0.7 (short dashed) and $f$=0 (dotted).
}
\label{fig5}
\end{figure}

For future experimental work we would like to propose to separate
$ R_{T'}$ and $R_{TL'}$. 
The sensitivity  of $ R_{T'}$ to $G_{M}^{(n)}$ is 
larger than for the asymmetry  $ A_{T'}$. 
This is demonstrated in  Fig.~\ref{fig5}
in comparison to  Fig.~\ref{fig3}
again for the FULL calculation.
Again we quantify that study by evaluating a $\chi ^2$, defined now as 

\begin{equation}
\chi^{2} ( R_{T'}, A_{T'})\, \equiv \,
\sum\limits_i
\frac{
\left( R_{T'}^{(i)} (A_{T'}^{(i)})  (f=1) 
-  
R_{T'}^{(i)} (A_{T'}^{(i)})  (f=1.3) \right) ^2
}
{
\left( R_{T'}^{(i)} (A_{T'}^{(i)})  (f=1) \right) ^2
} {\rm ,}
\label{eq67}
\end{equation}
where $i$ runs over the $\omega$-values, in which
we carried out the calculations.
We find $ \chi^{2} (R_{T'}) $ = 3.1 
and  $ \chi^{2} (A_{T'}) $ = 2.3.
Thus $R_{T'}$ has a stronger dependence on the magnetic neutron form factor 
(modified by the strength factor $f$ ) than $A_{T'}$. For the sake of curiosity
 Fig.\ref{fig5} also includes the results putting $G_M^{(n)}=0$ ($f$=0).

\section{ Summary }

Inclusive 
scattering of polarised electrons on polarised $^3$He has been evaluated 
taking the final state interaction fully into account. Realistic NN forces have been 
used and the 3N bound state and the 3N continuum are evaluated consistently solving 
the corresponding Faddeev equations. A formalism proposed in \cite{cite14}, which 
is ideal for inclusive processes and avoids the tedious direct integration 
 of over 
all final state configurations, has been generalized to handle new types of structure 
functions composed of different current components.

   The most simple picture of polarized $^3$He to be a polarized neutron target 
fails quantitatively for the energy and momentum transfers considered.
That picture relies on the assumption, that the principal S-state is by far 
dominant. This is not at all true for the transverse- longitudinal asymmetry 
$A_{TL'}$, which receives important contributions from the remaining pieces of  
the $^3$He wave function, but also for the transverse asymmetry $A_{T'}$, where 
the results change significantly when the principal S-state approximation is 
replaced by the full and correct $^3$He wave function. 

 We also find that the often used plane - wave impulse approximation ( here denoted 
 by PWIA' ) is insufficient. In PWIA' one takes the NN  force in the final state 
 into account for the pair of nucleons which are spectators to the single 
 nucleon photon absorption of the third nucleon. 
This is quite  insufficient for $A_{T'}$ and $A_{TL'}$. 
 The correct antisymmetrization of the final 3N continuum is important and 
 above all the final state interaction only all three nucleons (FULL calculation).

 In the FULL calculation the data for $A_{T'}$ can be described quite well 
 using the Gari-Kr\"umpelmann electromagnetic nucleon form factors. The dependence of 
 that observable $A_{T'}$ on the neutron $F_1$ form factor is weak and unimportant. 
 We optimized the choice of $G_M^{(n)}$ to the data, with the result that 
 the factor f=1 for the choice of Gari-Kr\"umpelmann parametrization was best.
 This appears to agree with preliminary results achieved in electron scattering 
 on the deuteron\cite{cite21}.
 
 In the case of $A_{TL'}$ the FULL calculation shows now the enhancement near the 3N 
 breakup threshold, which is present in the data and which was not provided by
  the plane wave impulse approximation used up to now. 

  For both observables $A_{T'}$ and $A_{TL'}$ more precise data would be very 
  welcome in order to probe the theoretical assumptions more stringently 
  and to extract more accurate information on $G_M^{(n)}$.

  A more thorough investigation of $A_{TL'}$ with respect to the contribution 
  of the proton and the $^3$He wave function component is planned. Because of 
  lack of computer time it could not be included in this study.

  We would also  like to  point out that data for $R_{T'}$ and $R_{TL'}$
  would be more sensitive to electromagnetic nucleon form factors than the 
  asymmetries. From the theoretical point of view mesonic exchange currents 
  should be added and the treatment of relativity remains a pending problem.

\acknowledgments
The authors are indebted to Dr. H. Gao and Dr. C. E. Jones for 
providing the details of their data.
This work was supported by the
Polish Committee for Scientific Research under Grant~PB~1031
and the Science and Technology Cooperation Germany-Poland
under Grant No.~XO81.91.
The numerical calculations have been per\-formed on the Cray T90 of the
H\"ochst\-leis\-tungs\-re\-chen\-zen\-trum in J\"ulich, Germany.


\section{Appendix}

The structure function $R_{T}$ has the form 

\begin{eqnarray}
R_{T} =\sum_{m' {\tau}' } \int df' \delta (M+\omega -P_{0}') ( \vert N_{1 }
\vert ^{2} + \vert N_{-1} \vert ^{2}) 
\end{eqnarray}
Using (\ref{eq11}) and (\ref{eq22}) one can write

\begin{eqnarray}
R_{T}&=&- { 3 \over \pi } \mbox{Im} \langle \Psi_{^{3}\mbox{He}} \vert j_{1} ^{(1)\dagger}
G_{0 } (1 + P) \vert U_{j_{1}} ^{(1)} \rangle 
      - { 3 \over \pi } \mbox{Im} \langle \Psi_{^{3}\mbox{He}} \vert j_{-1} ^{(1)\dagger}
G_{0 } (1 + P) \vert U_{j_{-1}}^{(1)} \rangle \cr
&=&
-{ {3 \over \pi }} \mbox{Im } 
[ 
{\sum \hskip -13 pt \int}
{ 1 \over { E + i \epsilon - {
      p^{2 }\over m } - { 3 \over {4 m } } q^{2 } } } \sum _{m'} \sum _{m''}
\cr
&{}&
D^{(1/2) *} _{m',m}  D^{(1/2)} _{m'',m} \langle p q \alpha \vert ( 1 + P ) j_{1}
^{(1)}  \vert \Psi _{^{3}\mbox{He }}  m' \rangle ^{* } 
\langle p q \alpha \vert U_{j_{1}}^{(1)}  m''\rangle ]  \cr
&{}&
-{ {3 \over \pi }} \mbox{Im } 
[ 
{\sum \hskip -13 pt \int}
{ 1 \over { E + i \epsilon - {
      p^{2 }\over m } - { 3 \over {4 m } } q^{2 } } } \sum _{m'} \sum _{m''}
\cr
&{}&
D^{(1/2) *} _{m',m}  D^{(1/2)} _{m'',m} \langle p q \alpha \vert ( 1 + P ) j_{-1}
^{(1)}  \vert \Psi _{^{3}\mbox{He }}  m' \rangle ^{* } 
\langle p q \alpha \vert U_{j_{-1} }^{(1)} m''\rangle ] 
\end{eqnarray}
Since $j_{1}^{(1)}$ ($j_{-1}^{(1)}$) increases (decreases) the $m$ magnetic
quantum by 1 one  gets
\begin{eqnarray}
R_{T}
&=&-{ {3 \over \pi }} \mbox{Im }  
{\sum \hskip -13 pt \int}
{ 1 \over { E + i \epsilon - {
      p^{2 }\over m } - { 3 \over {4 m } } q^{2 } } }
[ \vert D_{{1\over2} m} \vert ^{2} \langle p q \alpha {\cal J} {3\over 2} \vert (
1 + P ) j_{1}^{(1)} \vert \Psi_{^{3}\mbox{He }} {1\over2} \rangle ^{*} \langle
p q {\cal J} {3 \over 2 } \vert U_{j_{1}} ^{(1)} {1\over2} \rangle 
\cr
&+& \vert D_{-{1\over2} m} \vert ^{2} \langle p q \alpha {\cal J} {1\over 2} \vert (
1 + P ) j_{1}^{(1)} \vert \Psi_{^{3}\mbox{He }} -{1\over2} \rangle ^{*} \langle
p q {\cal J} {1 \over 2 } \vert U_{j_{1}} ^{(1)} -{1\over2} \rangle ]
\cr
&-&{ {3 \over \pi }} \mbox{Im }  
{\sum \hskip -13 pt \int}
{ 1 \over { E + i \epsilon - {
      p^{2 }\over m } - { 3 \over {4 m } } q^{2 } } }
[ \vert D_{{1\over2} m} \vert ^{2} \langle p q \alpha { \cal J} -{1\over 2} \vert (
1 + P ) j_{-1}^{(1)} \vert \Psi_{^{3}\mbox{He }} {1\over2} \rangle ^{*} \langle
p q {\cal J} -{1 \over 2 } \vert U_{j_{-1}} ^{(1)} {1\over2} \rangle 
\cr
&+& \vert D_{-{1\over2} m} \vert ^{2} \langle p q \alpha {\cal J} -{3\over 2} \vert (
1 + P ) j_{-1}^{(1)} \vert \Psi_{^{3}\mbox{He }} -{1\over2} \rangle ^{*} \langle
p q {\cal J} -{3 \over 2 } \vert U_{j_{-1}} ^{(1)} -{1\over2} \rangle ]
\end{eqnarray}
Using the phase relation (\ref{eq37}), in addition 
\begin{eqnarray}
\langle p q \alpha - {3\over 2} \vert ( 1 + P ) j _{-1}^{(1)} \vert \Psi
_{^{3}\mbox{He }} -{1\over2} \rangle = (-)^{{\cal J}-{1\over2}} \Pi < p q
\alpha {\cal J} {3\over 2} \vert (1 + P) j_{1}^{(1)} \vert \Psi_{^{3}\mbox{He
    }} {1\over2} \rangle 
\end{eqnarray}
and corresponding ones for 
$
\langle p q \alpha {\cal J} -{1 \over2} \vert U_{j_{-1}}^{(1)} {1\over 2}
\rangle $ and $
\langle p q \alpha {\cal J} -{3 \over2} \vert U_{j_{-1}}^{(1)} -{1\over2}
\rangle $ one arrives at (\ref{eq34}).
The structure function $R_{T'}$ has the form 

\begin{eqnarray}
R_{T'} =\sum_{m' {\tau}' } \int df' \delta (M+\omega -P_{0}') ( \vert N_{1 }
\vert ^{2} - \vert N_{-1} \vert ^{2}) 
\end{eqnarray}
A corresponding manipulation as above yields 

\begin{eqnarray}
R_{T'}&=&
-{ {3 \over \pi }} \mbox{Im }  
{\sum \hskip -13 pt \int}
{ 1 \over { E + i \epsilon - {
      p^{2 }\over m } - { 3 \over {4 m } } q^{2 } } }
\cr
&{}&
[ \langle p q \alpha {\cal J}
 {3\over 2} \vert ( 1 + P ) j_{1} ^{(1)} \vert \Psi
_{^{3}\mbox{He }} {1 \over 2} \rangle ^{*} \langle p q \alpha {\cal J}
{3 \over 2}
\vert U_{j_{1}} ^{(1)} {1\over2 } \rangle ( \vert D_{{1\over 2} m} \vert^{2 }
- \vert D_{-{1\over2} m} \vert ^{2} ) 
\cr 
&+&
< p q \alpha { \cal J} {1\over2} \vert ( 1 + P ) j_{1} ^{(1)} \vert \Psi
_{^{3}\mbox{He }} -{1\over2} \rangle ^{*} \langle p q \alpha {\cal J}
{1\over2}\vert U_{j_{1}}
^{(1)} -{1\over2} \rangle ( \vert D_{-{1\over2} m} \vert ^{2} -\vert
D_{{1\over2} m} \vert ^{2} ) ] 
\end{eqnarray}

Using (\ref{eq25}) one ends up with (\ref{eq40}).


\widetext

\begin{table}
\caption{
The experimental setup of Refs. \protect\cite{cite7,cite8,cite9}.
\label{table1}}
\begin{tabular}{lccccccccc}
& $k_0$ & $\Theta$  & $\theta_A$  & $\phi_A$ & $\omega$ & 
 $\omega_{QE}$ & $Q_{QE}$ & $\theta^*_{QE}$ & $\phi^*_{QE}$\\
& (MeV) & (\cdeg) & (\cdeg) & (\cdeg) & (MeV) & (MeV) & (MeV/c) & 
 (\cdeg) & (\cdeg) \\
\hline
Ref. \protect\cite{cite7} & 370 & 91.4 & 42.5 & 180 & 91 -- 150 & 
  107 & 460 & 8.9 & 180 \\
Ref. \protect\cite{cite8} & 370 & 70.1 & 42.5 &   0 & 73 --  97 &  
  76 & 386 & 88.1 & 0\\
Ref. \protect\cite{cite9} & 370 & 70.1 & 42.5 &   0 & 40 --  52 &  
  76 & 386 & 88.1 & 0\\
\end{tabular}
\end{table}


\begin{references}

\bibitem{cite1}
R. G. Arnold \etal, Phys. Rev. Lett. {\bf 57} (1986) 174;
P. E. Bosted \etal, Phys. Rev. Lett. {\bf 68} (1992) 3841.
\bibitem{cite2}
P.~Markowitz \etal,  Phys. Rev. {\bf C48} (1993) R5;
J.~Jourdan, AIP Conference Proceedings 334,
Few-Body Problems in Physics, ed. F.~Gross,
Williamsburg 1994, page 339;
Proceeding of the 14th International Conference on Particles and Nuclei, 
eds. G. E. Carlson, J.J. Domingo, Williamsburg 1996 (World Scientific 1997), 
page 262.
\bibitem{cite3}
B.~Blankleider, R.~M.~Woloshyn,  Phys. Rev. {\bf C29} (1984) 538.
\bibitem{cite4}
J. L. Friar, B.~F.~Gibson, G.~L.~Payne,
A.~M.~Bernstein, T.~E.~Chupp, Phys. Rev. {\bf C42} (1990) 2310.
\bibitem{cite5}
C. E. Woodward \etal, Phys. Rev. Lett. {\bf 65} (1990) 698;
C. E. Jones-Woodward \etal,  Phys. Rev. {\bf C44} (1991) R571;
C. E. Jones \etal,  Phys. Rev. {\bf C47} (1993) 110.
\bibitem{cite6}
A. K. Thompson \etal,  Phys. Rev. Lett. {\bf 68} (1992) 2901.


\bibitem{cite7}
H. Gao \etal,  Phys. Rev. {\bf C50} (1994) R546.
\bibitem{cite8}
J.-O. Hansen \etal,  Phys. Rev. Lett. {\bf 74} (1995) 654.
\bibitem{cite9}
C. E. Jones \etal, Phys. Rev. C {\bf 52} (1995) 1520.

\bibitem{cite10}
R.-W. Schulze and P. U. Sauer, Phys. Rev. {\bf C48} (1993) 38.
\bibitem{cite11}
C. Ciofi degli Atti, E. Pace and G. Salm\'e, 
Phys. Rev. {\bf C51} (1995) 1108.

\bibitem{cite12}
H. Kamada, W. Gl\"ockle, J. Golak, Nuovo Cimento {\bf 105A} (1992) 1435.
\bibitem{cite13}
S. Ishikawa, H. Kamada, W. Gl\"ockle, J. Golak, H. Wita{\l}a,
Nuovo Cimento {\bf 107A} (1994) 305.
\bibitem{cite14}
S. Ishikawa, H. Kamada, W. Gl\"ockle, J. Golak and H. Wita\l a,
Phys. Lett. {\bf B339} (1994) 293.
\bibitem{cite15}
J. Golak, H. Kamada, H. Wita\l a, W. Gl\"ockle, 
S. Ishikawa, Phys. Rev. {\bf C51} (1995) 1638.
\bibitem{cite16}
J. Golak, H. Wita\l a, H. Kamada, D. H\"uber, S. Ishikawa,
W. Gl\"ockle, Phys. Rev. {\bf C52} (1995) 1216.

\bibitem{cite17}
T. W. Donnelly and A. S. Raskin,
Ann. Phys. (N.Y.) {\bf 169} (1986) 247.

\bibitem{cite18}
W. Gl\"ockle, {\em The Quantum Mechanical 
Few-Body Problem} (Springer-Verlag, Berlin Heidelberg 
New York Tokyo 1983).

\bibitem{cite19}
R. Machleidt, Advances in Nucl. Phys. {\bf 19} (1989) 189.

\bibitem{cite20}
M. Gari and W. Kr\"umpelmann, Phys. Lett. {\bf 173B} (1986) 10.

\bibitem{cite21}
J. Jourdan (private communication)
\end{references}
\end{document}